\documentstyle[psfig]{article}

\def\baselinestretch{2.}
\begin{document}

\title{ASCA Observation of the Crab-Like Supernova Remnant 3C58}

\author{
Ken'ichi {\sc Torii,$^{1}$} Patrick O. {\sc Slane,$^2$}\\ Kenzo {\sc Kinugasa,$^{3}$} Kiyoshi {\sc Hashimotodani,$^4$} and Hiroshi {\sc Tsunemi$^{4,5,1}$}\\
\\
{\it $^1$ NASDA TKSC SURP, 2-1-1 Sengen, Tsukuba, Ibaraki, 305-8505, Japan}\\
{\it E-mail (KT) torii.kenichi@nasda.go.jp}\\ 
{\it $^2$Harvard-Smithsonian Center for Astrophysics, 60 Garden Street
Cambridge, MA 02138, USA}\\
{\it $^3$ Gunma Astronomical Observatory,}\\
{\it 6860-86 Nakayama, Takayama, Agatsuma,Gunma, 377-0702, Japan}\\
{\it $^4$ Department of Earth and Space Science, Graduate School of
Science, Osaka University,}\\ 
{\it 1-1, Machikaneyama-cho, Toyonaka, Osaka, 560-0043, Japan}\\
{\it $^5$CREST, Japan Science and Technology Corporation (JST)}\\
}

\maketitle

\begin{abstract}

	We present here the X-ray observation of a Crab-like supernova
remnant (SNR) 3C58 with ASCA. We find that the integrated energy
spectrum over the nebula is consistent with previous results, showing
a power-law spectrum with the photon index $\gamma = 2.2-2.4$ modified
by interstellar absorption of about $(3-4)\times 10^{21}{\rm
cm^{-2}}$. Inclusion of a blackbody component which is attributable to
the central compact source significantly improves the spectral fit. Stringent
upper limits for any line emitting thin hot plasma are established. We
find for the first time that the nebular spectrum is harder in the
central part of the SNR, becoming softer toward the periphery, 
while the absorption column is uniform across the
nebula. Correspondingly, the nebular size decreases with increasing
photon energy which is a steeper function of radius than that of the
Crab nebula. The results are compared with synchrotron energy loss
models and the nature of the putative pulsar is discussed. Timing
analysis was performed to search for  pulsed X-ray emission from the
central compact source. No significant pulsations are observed, and we
present the upper limit for the pulsed fraction.

\end{abstract}

\section{Introduction}

	The source  3C58 (G130.7+3.1) is believed to be the remnant of
a historical supernova (SN) event in A.D. 1181 (e.g., Panagia, Weiler
1980). It shows a center-filled morphology both in the radio (e.g.,
Reynolds, Aller 1988) and X-ray bands (e.g., Becker et
al. 1982). X-ray observations reveal a point-like source in the midst of 
the nebula (Becker et al. 1982) which contributes 7\% of the total flux in
the ROSAT HRI energy band (0.1-2.4 keV) (Helfand et al. 1995). The
radio spectrum of the nebula is a power-law with an energy index of
$0.1$ (Green 1986) and the X-ray spectrum is a power-law with an
energy index of $1.19\pm 0.12$ (Asaoka, Koyama 1990). Though the
central object itself is radio quiet, a wisp-like structure elongated
in the north-south direction lies only $2.6$ arcseconds
west of the X-ray compact source (Frail, Moffett 1993). Similar
wisp-like structures in the vicinity of the central objects are often
found in the Crab-like SNRs, not only in the radio band
(Bietenholz et al. 1991; Frail, Moffett 1993) but also in the optical
band (Hester et al. 1995). These structures are believed to be
associated with the termination of the relativistic pulsar wind by the
ambient pressure. These characteristics, along  with the nonthermal nature of
the extended emission in 3C58, imply that the central source is a
neutron star even though no coherent pulsations have been detected at
any wavelength to date. From these characteristics, 3C58 is classified
as a Crab-like SNR or a {\it plerion} (Weiler, Panagia 1978).

	Though the Crab nebula and 3C58 have much in common, there are
considerable differences as well. The radio and X-ray luminosities for
3C58 are factors of $\sim 10$ and $\sim 2000$, respectively, below
those of the Crab nebula (Helfand et al. 1995). As these objects are
powered by the rotational energy loss of the central neutron star, the
differences found in the surrounding nebulae should give insights into
the properties of the young neutron stars.

	The modeling of the Crab-like synchrotron nebulae has evolved
with a strong bias on the Crab nebula itself due to its outstanding
brightness. Models have been developed to explain the multi-wavelength
spectrum and its evolution (e.g., Pacini, Salvati 1973; Reynolds,
Chevalier 1984; Reynolds, Chanan 1984; Bandiera et al. 1996), the flow
of relativistic particles and the corresponding morphology of the
nebula (e.g., Rees, Gunn 1974; Aschenbach, Brinkmann 1975; Kennel,
Coroniti 1984a, b; Chiueh, Li, Begelman 1998). Following the model of Rees and Gunn (1974),
Kennel and Coroniti (1984a, b hereafter KCa, b) proposed a
magnetohydrodynamical (MHD) flow model with a standing shock having an
extremely low magnetization parameter, $\sigma$ (the ratio of the
Poynting to particle energy fluxes in the upstream wind). This model
successfully explains the previously observed variation in
nebular size with energy from the optical to the hard
X-ray band under the suitable boundary conditions.

	The distance to 3C58 was derived by H\,{\footnotesize I}
observations by several authors (Goss et al. 1973; Williams 1973;
Green, Gull 1982; Roberts et al. 1993). Following the discussions made
by Roberts et al. (1993), we adopt 3.2 kpc throughout this paper.

	In this paper, we report the X-ray observation of 3C58 with
the spectro-imaging detectors on board ASCA in the energy range 0.5--10
keV.

\section{Observation and Data Reduction}

	3C58 was observed by ASCA (Tanaka et al. 1994)
from 12 to 14 in September, 1995. As well as a proposed target of the
AO3 observing phase, it was observed as an SIS (Solid-state Imaging
Spectrometer) (Burke et al. 1994) calibration source. These
observations were carried out with four different satellite attitudes. 
The fourth observation was made to obtain the accurate background in
the adjacent region of the object. The SIS was operated in
1CCD mode with a read-out cycle of 4 s. The Gas Imaging Spectrometer (GIS),
a position sensitive gas scintillation proportional counter
(Ohashi et al. 1996; Makishima et al. 1996), was operated in PH
mode. To search coherent pulsations, portions of the observation 
were carried out in a mode which allocates more bits to timing resolution 
than for standard observations. This sacrifices some spatial and
spectral resolution. The time resolution of the GIS for the first
observation was 1/16 s for high bit rate and 1/2 s for medium bit
rate. For the other observations, the resolution was 1/1024 s and 1/128
s for high and medium bit rate, respectively. The observation date and
the observation mode are summarized in table
\ref{observation_date_mode}.

	The raw data were screened with the standard screening
parameters. The screening criteria restrict events to those obtained
outside the South Atlantic Anomaly (SAA), for which the cut-off
rigidity is $> 6$~GeV/c, and for which the angle of the source from
the Earth rim is restricted to $> 5$~ deg for the GIS, $> 10$~deg
(25~deg) for the SIS for the dark (bright) Earth rim angle. For the
SIS, flickering pixels are removed, only grades 0, 2, 3, and 4 are
selected, and four readout cycles are removed after the satellite
passage of the SAA as well as after the satellite passage of the
day-night transition.  The effective exposure time after the data
reduction is summarized in table \ref{exposure} for each observation.

\section{Results}

\subsection{Energy Spectrum of the Whole Remnant}

\begin{figure}[p]
\psfig{figure=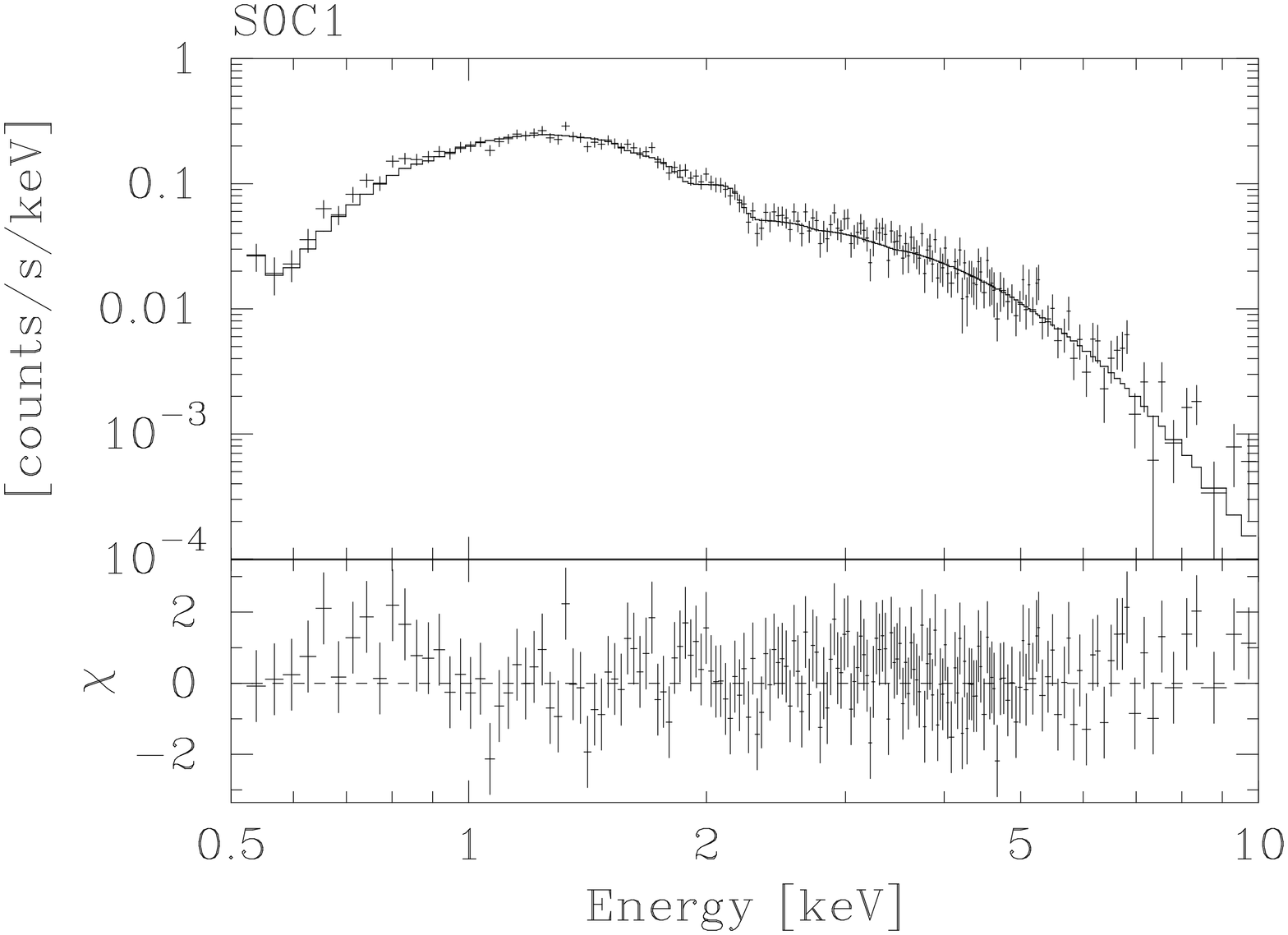,height=8.0cm,angle=270}
\psfig{figure=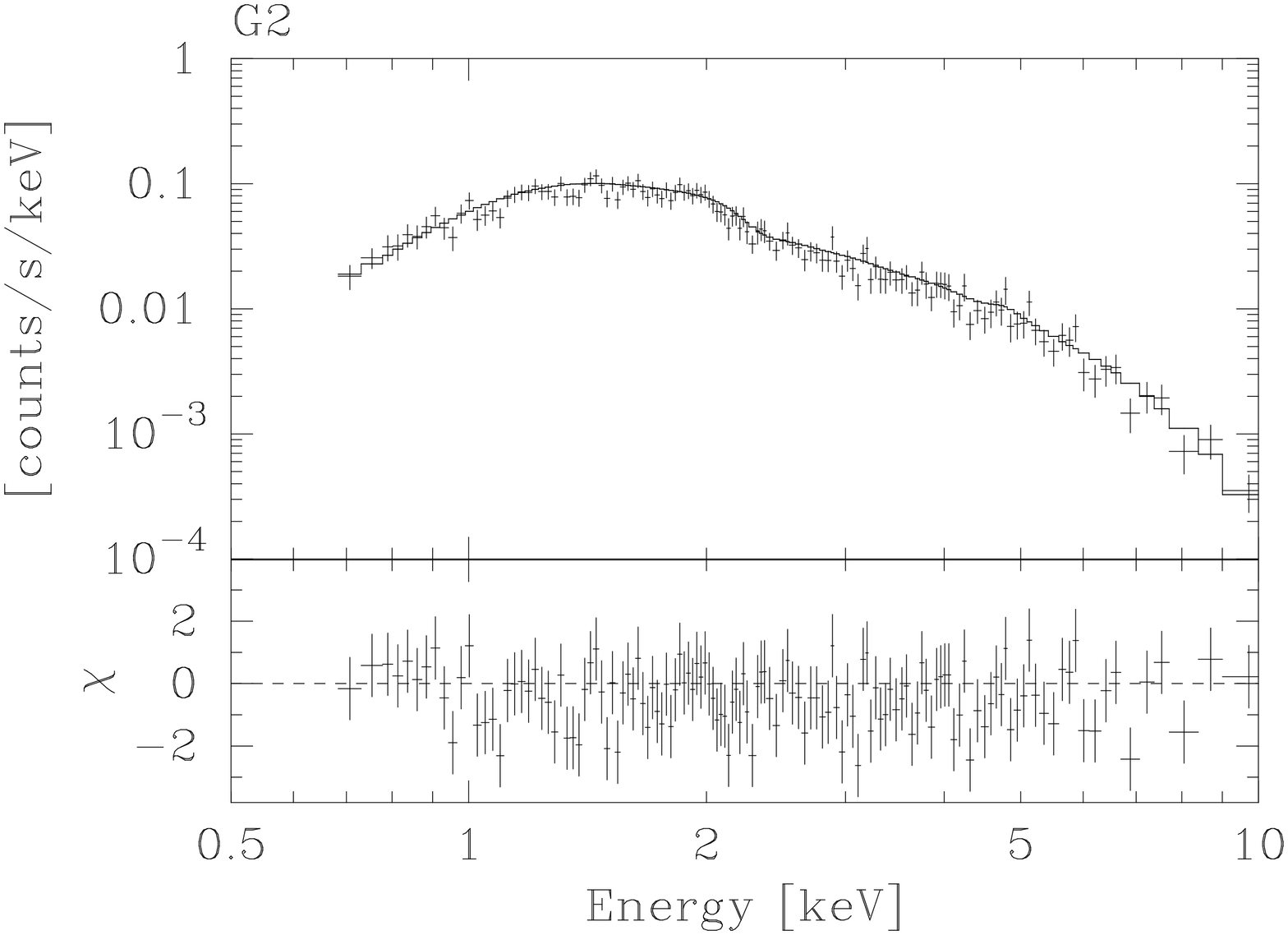,height=8.0cm,angle=270}
\caption{}
\label{representative_spectra}
\end{figure}

	We first investigated the energy spectrum of the whole SNR,
including the compact source. To check the consistency between the
detectors, energy spectra were extracted for each detector of each
sensor using photons within 4 arcminutes from the emission center.
The spectra were fitted with a power-law function modified by the
interstellar absorption (Morrison, McCammon 1983). A formally acceptable
fit was obtained for each data set by using this model.  Since we found
no significant systematic deviation above the level of the current
status of the detector calibration of ASCA, we fitted the spectra from
each observation simultaneously to improve the statistics. We show
representative spectra in figure \ref{representative_spectra} and the
best-fit spectral parameters are summarized in table
\ref{all_spectra_parameters}.

	Becker et al. (1992) obtained
$\gamma=1.5\stackrel{+0.5}{_{-0.2}}$ and $N_{{\rm
H}}=(1.8\pm0.5)\times10^{21}{\rm cm^{-2}}$ with the Einstein SSS, but
a somewhat steeper power-law index ($\gamma = 2$) and approximately
twice the column density using the IPC.

By fixing the absorption column to 
$N_{{\rm H}}=1.8\times10^{21}{\rm cm^{-2}}$, the value
determined with the Einstein SSS, Davelaar et al. (1986)
obtained $\gamma = 2.30\pm0.26$ with the EXOSAT ME. Asaoka and
Koyama (1990) obtained $\gamma=2.19\pm0.12$ and $N_{\rm H} < 10^{21.5}
{\rm cm^{-2}}$ with the Ginga LAC. Helfand et al. (1995) derived
$\gamma = 1.70\pm0.30$ and $N_{\rm H} = (3.25\pm0.7)\times 10^{21}
{\rm cm^{-2}}$ for the nebula by considering the Einstein SSS, IPC,
ROSAT HRI, and the 21 cm (Green, Gull 1982) measurements, and a
blackbody of $3.5\times 10^6 \;{\rm K}$ for the central compact
source.
Taking into account the statistical and systematic errors of the data
in the literature, these values are in reasonable agreement with
those summarized in table 3.
By using the first set of the
SIS data, the observed flux is $1.3\times 10^{-11}{\rm ergs\cdot
s^{-1}\cdot cm^{-2}}$ and the unabsorbed flux is $1.9\times
10^{-11}{\rm ergs\cdot s^{-1}\cdot cm^{-2}}$ in the range 0.5--10
keV. The statistical error in the flux determination is about 4\% and
the systematic error for the absolute flux is about 10\%. The
corresponding luminosity is $2.4\times10^{34} d_{{\rm 3.2kpc}}^2
{\rm ergs \cdot s^{-1}}$ in the range 0.5--10 keV.

	Though we see no line structure in the raw energy spectra by
eye, we might expect some amount of thermal plasma of temperature
$\sim10^7 {\rm K}$ that is shock-heated at the outgoing blast wave, and
possibly at the reverse shock. To search for evidence of such thermal
emission, we performed the following analysis to establish limits on
the emission measure (EM $= \int n_{{\rm e}} n_{{\rm H}} dV$, where
$n_{{\rm e}}$, $n_{{\rm H}}$, and $V$ denote the electron number density, the
hydrogen number density, and the volume, respectively) by using a thin
thermal plasma emission code.

\begin{figure}[p]
\psfig{figure=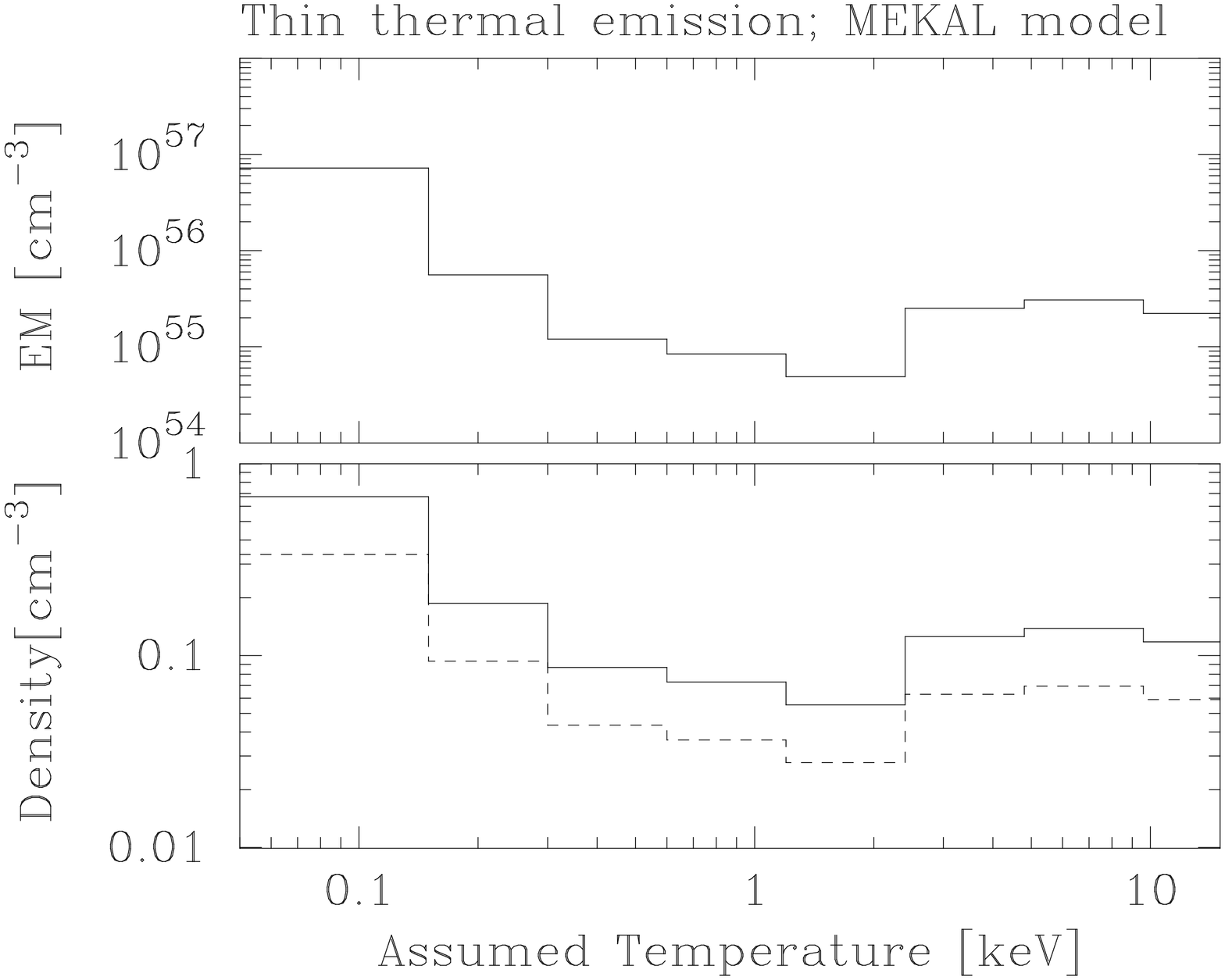,height=10.0cm,angle=270}
\caption{}
\label{thermal_mekal}
\end{figure}

	We used the MEKAL model in XSPEC ver.9.0 which calculates the
emission from an optically thin thermal plasma in collisional
ionization equilibrium. The model includes line emission from heavy
elements based on the model calculations of Mewe and Kaastra (e.g.,
Mewe et al. 1985) with the Fe L calculations by Liedahl et al. (1995). 
We assumed heavy elemental abundances equal to the solar values of
Anders and Grevesse (1989). All the SIS and GIS data were used
simultaneously to obtain the stringent upper limit. If both the
power-law index and the absorption are allowed to vary, along with the
thermal component, we find large absorption and a large power-law
index which is incompatible with the Ginga result at higher energy
(Asaoka, Koyama 1990). Therefore, we fix the power-law index and the
absorption to be the best-fit values obtained without the thermal
component. The resultant upper limit of the EM is shown in the upper
panel of figure \ref{thermal_mekal} for the assumed temperature range
0.1--12.8 keV. The corresponding upper limit of the rms density of the
thermal plasma is shown in the lower panel assuming a spherical volume whose
angular radius is 4 arcminutes. The solid line shows the limit when
the filling factor is 1/4 which corresponds to the Sedov type self
similar solution (Sedov 1993) and the dashed line shows the limit for
a uniform sphere and the filling factor of unity.

	Another possible thermal component is blackbody radiation from
the cooling neutron star surface. A blackbody temperature of about $10^6$K
from the entire surface, or an even higher temperature due to polar
cap heating, may be expected (e.g., Yancopoulos et al. 1994; Slane
1994; Slane, Lloyd 1995; Helfand et al. 1995). Helfand et al. (1995)
derived a crude constraint on the spectral shape of the compact source
based on the flux ratio between the Einstein HRI and the ROSAT
HRI. They obtained a blackbody temperature of $3.5\times10^6 {\rm K}$
and an emitting area of
$\sim 4.5\times 10^{10} d_{{\rm 3.2kpc}}^{2} \;{\rm cm^2}$. To examine
this component, we introduced a blackbody component to the
simultaneous fit of all the SIS and GIS spectra. Without the blackbody
component, the photon index, normalization of the power-law component,
and the absorption column are $\gamma=2.29\pm0.03$,
$N=(4.8\pm0.2)\times10^{-3}$ ${\rm photons\cdot s^{-1} \cdot
keV^{-1}\cdot cm^{-2}}$ at 1 keV, and $n_{{\rm
H}}=(4.0\pm0.2)\times10^{21}{\rm cm^{-2}}$ with $\chi^2/{\rm d.o.f.} =
1487/1330$. With the blackbody component, $\chi^2/{\rm d.o.f.} =
1474/1328$ which shows that the blackbody component is significant at
99.7\% level according to F-test. The best-fit spectral parameters are
$T_{{\rm blackbody}} = (5.1\stackrel{+0.6}{_{-0.5}})\times 10^6 \;{\rm
K}$ and the emitting 
area of $(1.2\stackrel{+0.8}{_{-0.7}})\times 10^{10} d_{{\rm
3.2kpc}}^{2} \; {\rm cm^2}$. The parameters for the power-law
component and the absorption column are $\gamma = 2.1\pm0.1$,
$N = (3.6\pm0.6)\times 10^{-3}\;{\rm photons \cdot s^{-1} \cdot keV^{-1}
\cdot cm ^{-2} }$ at 1 keV, and $n_{{\rm H}} = (3.3\pm0.4)\times
10^{21} {\rm cm^{-2}}$. With this model, the observed flux from the
power-law and the blackbody component is $1.1\times10^{-11}$ and
$8.4\times 10^{-13}{\rm ergs\cdot s^{-1}\cdot cm^{-2}}$ in the 0.5--10
keV range, respectively. The unabsorbed flux is $1.6\times10^{-11}$
and $1.2\times 10^{-12}{\rm ergs\cdot s^{-1}\cdot cm^{-2}}$,
respectively. The corresponding luminosity is $L_{{\rm
X,pow}}=2.0\times 10^{34}d_{{\rm 3.2kpc}}^2 {\rm ergs\cdot s^{-1}}$
and $L_{{\rm X,blackbody}}=1.5\times 10^{33} d_{{\rm 3.2kpc}}^2{\rm
ergs\cdot s^{-1}}$. We find the blackbody temperature derived here is
$\sim 1.5$ times higher than that of Helfand et al. (1995) and the
emitting area is $\sim 3.8$ times smaller. From the flux point of
view, however, these two results are remarkably consistent within
$\sim 20\%$. Thus we have confirmed the presence of a relatively high
temperature blackbody component (Helfand et al. 1995) in the spectrum
of the whole SNR. Due to the limited spatial resolution of ASCA and
the limited spectral resolution of Einstein HRI and Rosat HRI, the
temperature of the blackbody component can not be strictly
constrained. The contribution from the blackbody component is further
examined in the following spatially resolved analyses to confirm that
it predominantly comes from the central compact source.

	The power-law component of the X-ray spectrum determined with
the power-law plus blackbody model is shown in figure
\ref{multi_wavelength} together with the radio (Green 1986; Salter et
al. 1989 and the references therein), infrared (Green, Scheuer 1992;
Green 1994), and $\gamma$-ray (Fichtel et al. 1994) data. The infrared
and $\gamma$-ray data are the upper limits. The multi-wavelength
spectrum of the Crab nebula is also shown in the figure for comparison
(e.g., Saken et al. 1992; Wu 1981 and the references therein).

\subsection{Energy Spectra of Concentric Annular Regions}

	The synchrotron lifetime of the electrons producing the X-ray
emission varies as $\propto E^{-1}$. We thus expect a radial variation
in the spectral index of the nebula with emission in the inner regions
(where electron injection is most recent) exhibiting a flatter
spectrum than for the outer regions. To investigate such a scenario,
we performed spatially resolved spectral analysis. As the SIS has
better spatial resolution than the GIS, we used only the SIS data for
this analysis. We extracted energy spectra from concentric annular
regions centered on the emission peak. The width of each annulus was 1
arcminute and the outer radii were 1, 2, 3, and 4 arcminutes. We first
used a power-law function modified by the interstellar absorption to
examine the systematic effect in the procedure and then introduced a
blackbody component to the fit.

	Since the point spread function (PSF) of ASCA X-ray telescope
(XRT) (Serlemitsos et al. 1995) has a half power diameter of about 3
arcminutes, comparable in size with the 3C58 extended nebula, we
conducted the analysis in the following way to avoid the possible
systematic effects. We first fitted the 3C58 data with all the model
parameters free. We found that the power-law index was the
flattest in the central region, steepening toward the periphery,
while the absorption column was uniform across the nebula. To see the
possible systematic effects, we applied the same procedure to the
archival data of 3C273 (sequence number 71038020), a bright point
source. We found neither the systematic steepening of the power-law
index nor any systematic change of the absorption column depending on
the radius. Since we found no significant systematic effects due to
the instrumental response functions, we fitted the 3C58 data with the
absorption column and the blackbody temperature (when included) fixed
at the best-fit value obtained for the whole remnant to obtain
statistically stringent limit on the other spectral parameters.

	The resultant power-law indices are shown in figure
\ref{annular_gamma} with 90\% confidence errors. Squares show the
result without the blackbody component and the circles show the result
with the blackbody component. We find significant softening of the
energy spectrum toward the periphery of the nebula, as expected from
the synchrotron energy loss of the relativistic particles. Also, when
the blackbody component is included, we find that the fraction of the
flux from the blackbody component is about 10\% in the inner part
while it becomes negligibly weak at the outermost annulus ($9.4\%$,
$9.7\%$, $4.0\%$, and $<0.6\%$ in the 0.5--10 keV range from the
innermost toward the outermost annulus). This supports the
interpretation that the blackbody radiation comes from the central
source in 3C58. Thus we have shown that the spectrum consists of two
components, one the spatially extended power-law component and the
other a blackbody component at the center, and the former shows
softening toward the periphery of the nebula. Since the spectral
indices derived here are obtained after the convolution by the XRT
PSF, we further quantify this softening in the following section in
terms of the energy-size relation.

\subsection{Image Fitting Analysis}

	In the method we adopted in the previous section, the spectrum
for a given annulus is contaminated by the photons originating from
the adjacent annuli due to the wings of the XRT PSF.  In this section,
we explore the spatial distribution of the energy spectrum using a
different approach. We have extracted images in several energy ranges
and fitted them with some appropriate function convolved by the
PSF. In this method, the extracted image of each energy range does not
suffer much from the photons whose energy is out of the energy range
considered. This is justified for the SIS data due to its high energy
resolution. We used subroutines and the standard calibration files
provided by the instrumental team as part of the {\it jbldarf} and
{\it ftools}
packages to calculate the PSF. The azimuthal structures due to the
telescope quadrants are not taken into account in the PSFs in these
functions.

	Since we confirmed the blackbody component in the spectrum
which is attributable to the compact source at the center of the
nebula, we utilize a model function which comprises the nebula
component, the compact source component, and the background component. 
We employ broad and narrow Gaussians for the compact source and the
nebula, respectively. The model function is expressed as follows.
\begin{equation}
f(x,y) = A \cdot {\rm exp}(-\frac{(x-x_0)^2}{2\sigma_{1x}^2} -
\frac{(y-y_0)^2}{2\sigma_{1y}^2}) + B \cdot {\rm
exp}(-\frac{(x-x_0)^2}{2\sigma_{2x}^2} - \frac{(y-y_0)^2}{2\sigma_{2y}^2}) + C
\label{gaussian_delta_2d}
\end{equation}
where $A$, $B$, and $C$ are the normalizations for the nebula, the
compact source, and the background component, respectively; $x$ and
$y$ are the positions in the right ascension and declination
coordinate. Since the major axis of the nebular emission is well
aligned in the east-west direction, we did not rotate the function
around the center. The size of the nebula is characterized by
$\sigma_{1x}$ and $\sigma_{1y}$.
 $\sigma_{2x}$ and $\sigma_{2y}$ are
fixed at 5 arcseconds which we find are appropriate to reproduce a
compact source at the center. Since the contribution from the
blackbody component to the observed flux is, based on the best fit
spectral parameters, $7.9\%$, $15.6\%$, $16.7\%$, and $2.0\%$ in the
0.5--1, 1--1.5, 1.5--3, and 3--10 keV range respectively, the ratio
between A and B is fixed for each energy range to reproduce these
fractions.

	To examine the systematic error of the XRT PSF, we fitted the
3C58 images by using the 3C273 image as a PSF. A similar energy-size
relation was obtained, but the derived sizes were different by +4\%,
+11\%, +14\%, and +11\% in the EW direction and -6\%, -8\%, -11\%, and
-20\% in the NS direction, in the 0.5--1, 1--1.5, 1.5--3, and 3--10
keV range, respectively. Thus we found at most 20\% systematic error
between the analyses made with the two kinds of PSFs. Since these
systematic errors dominate the statistical errors, we quote 20\% as
the total error.

	Figure \ref{radial_profile} shows the azimuthally averaged
radial profile of the data, the convolved model function, and the PSF.
Figure \ref{xy} shows the position of the Gaussian center.  Error bars
show the 90\% confidence positional uncertainty of ASCA which is 40
arcseconds (Gotthelf 1996; Gotthelf, Ishibashi 1997). We noted that
the central position of the concentric annular regions used in the
previous analyses agreed with the Gaussian centers within $\sim0.1$
arcminutes. The absolute value of the Gaussian center is displaced
about 40 arcseconds to the east compared with the position of the
compact source determined by ROSAT HRI (Helfand et al. 1995), an
offset which is consistent with the systematic positional uncertainty.

	Figure \ref{sigma} shows the size of the nebula as a function
of the observing photon energy. The nebular size is found to show a
monotonic decrease toward higher energy. A power-law fit for the nebular
size as a function of the energy yields
\begin{equation}
r_{{\rm FWHM}}{\rm [arcsec]} = (190\pm 40) \times E{\rm [keV]} ^{-(0.5\pm 0.2)}
\label{fwhm_delta_ew}
\end{equation}
in the east-west direction and
\begin{equation}
r_{{\rm FWHM}}{\rm [arcsec]} = (140 \pm 30) \times E{\rm [keV]} ^{-(0.5\pm 0.2)}
\label{fwhm_delta_ns}
\end{equation}
in the north-south direction.

\subsection{Timing Analysis}

	Though the emission characteristics of 3C58 strongly indicate
the presence of a central pulsar, to date no coherent pulsations from
the central object have been reported (Helfand et al. 1995). To search
for a periodic signal from the central object, we performed a power
spectral density (PSD) analysis on the GIS data. Photons extracted
within 4 arcminutes from the emission center were used in the
analysis. Barycentric corrections were made to the photon arrival
times and a fast Fourier transformation (FFT) was applied to obtain the
PSD. To improve the statistics, GIS2 and GIS3 data were co-added and
they were analyzed by using a software developed by ourselves (e.g.,
Torii et al. 1997).

	We found no significant pulsations at the 99\%
confidence level in the frequency range 0.01--64 Hz. The higher
frequency range was limited by the time resolution of medium bit rate
data. The upper limits for the pulsed fraction for a sinusoidal pulse
shape were calculated according to the method described by Vaughan et
al. (1994). The resultant 99\% confidence upper limits for the entire
selected region are summarized in table \ref{upper_limit_new} for the
three energy ranges, 0.7--10, 0.7--2, and 2--10 keV. 

Since the number of photons is dominated by those originating from the
nebula, with only $\sim 10\%$ originating from the central compact
source, the pulsed fraction of the compact source is not tightly
constrained. For the whole energy band (0.7--10 keV), the fraction of
photons from the blackbody component expected from the best-fit
spectral parameters is 11.7\%. Therefore, if the blackbody component
is pulsating, the upper limits for the pulsed fraction of the compact
source are estimated as 
62\%, 82\%, 80\%, and 74\% 
in the 0.01--1, 1--8, 8--32, and 32--64 Hz range, respectively.

\section{Discussion}

\subsection{Multi-Wavelength Spectrum}

\begin{figure}[p]
\psfig{figure=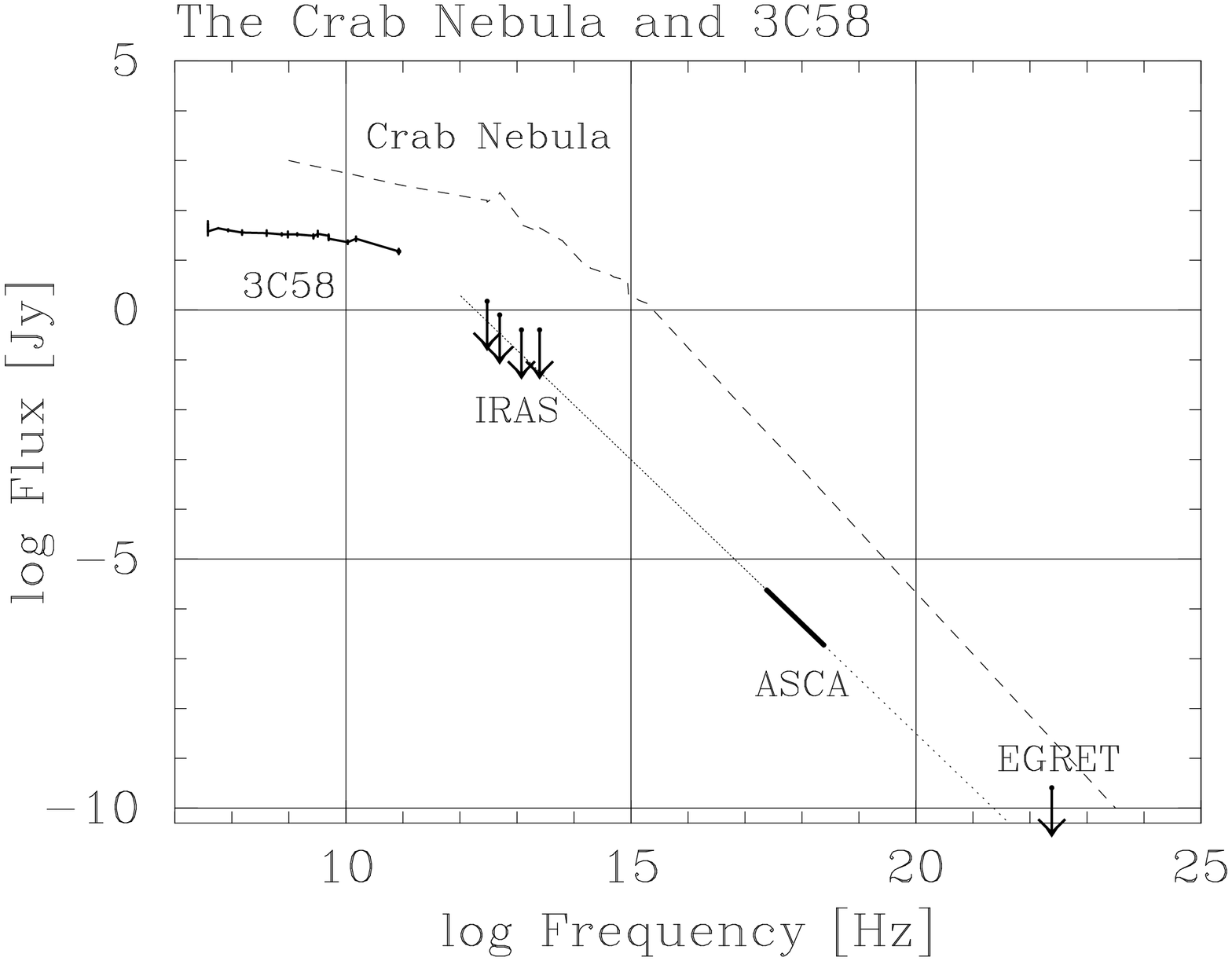,height=10.0cm,angle=270}
\caption{}
\label{multi_wavelength}
\end{figure}

\begin{figure}[p]
\psfig{figure=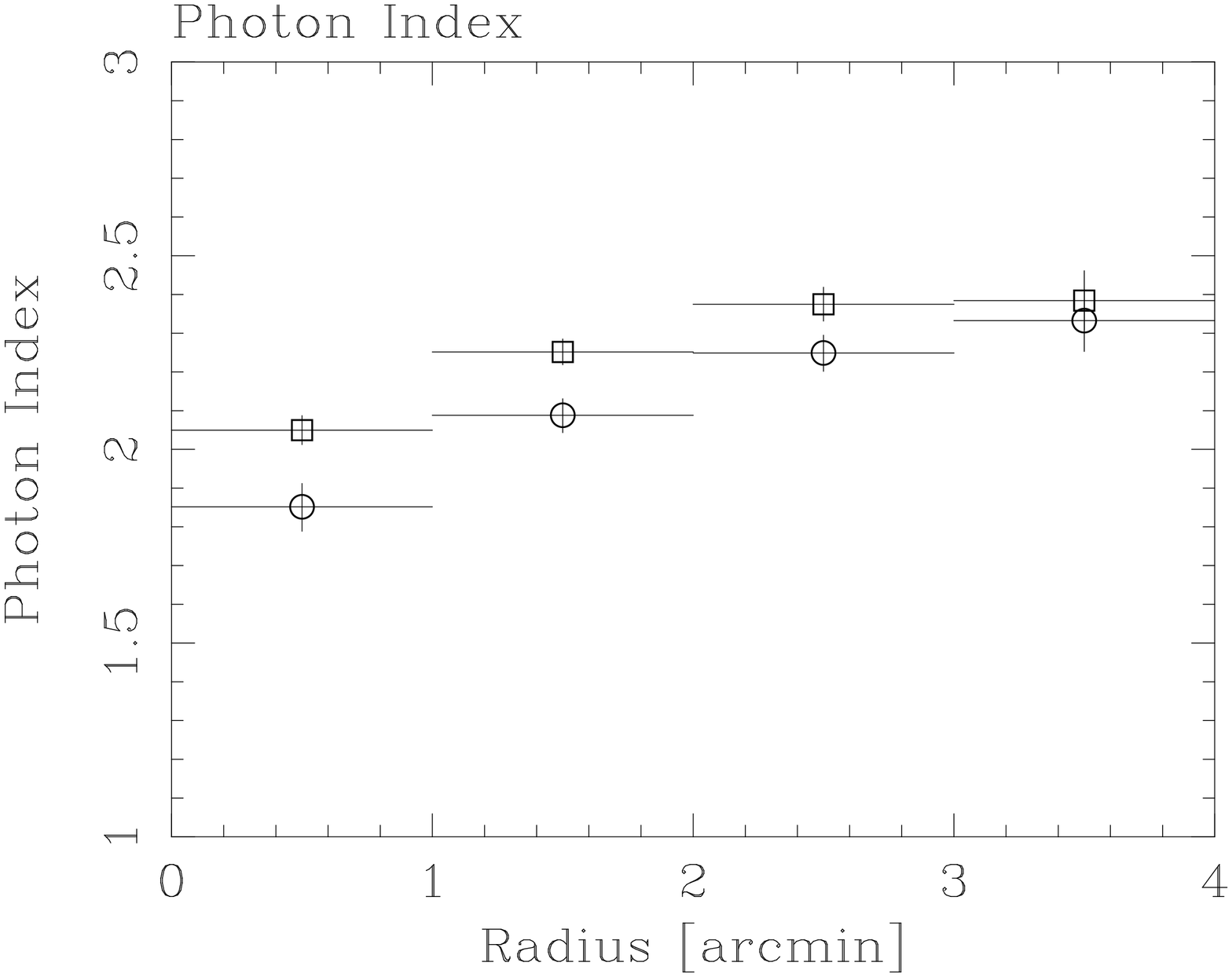,height=10.0cm,angle=270}
\caption{}
\label{annular_gamma}
\end{figure}

\begin{figure}[p]
\psfig{figure=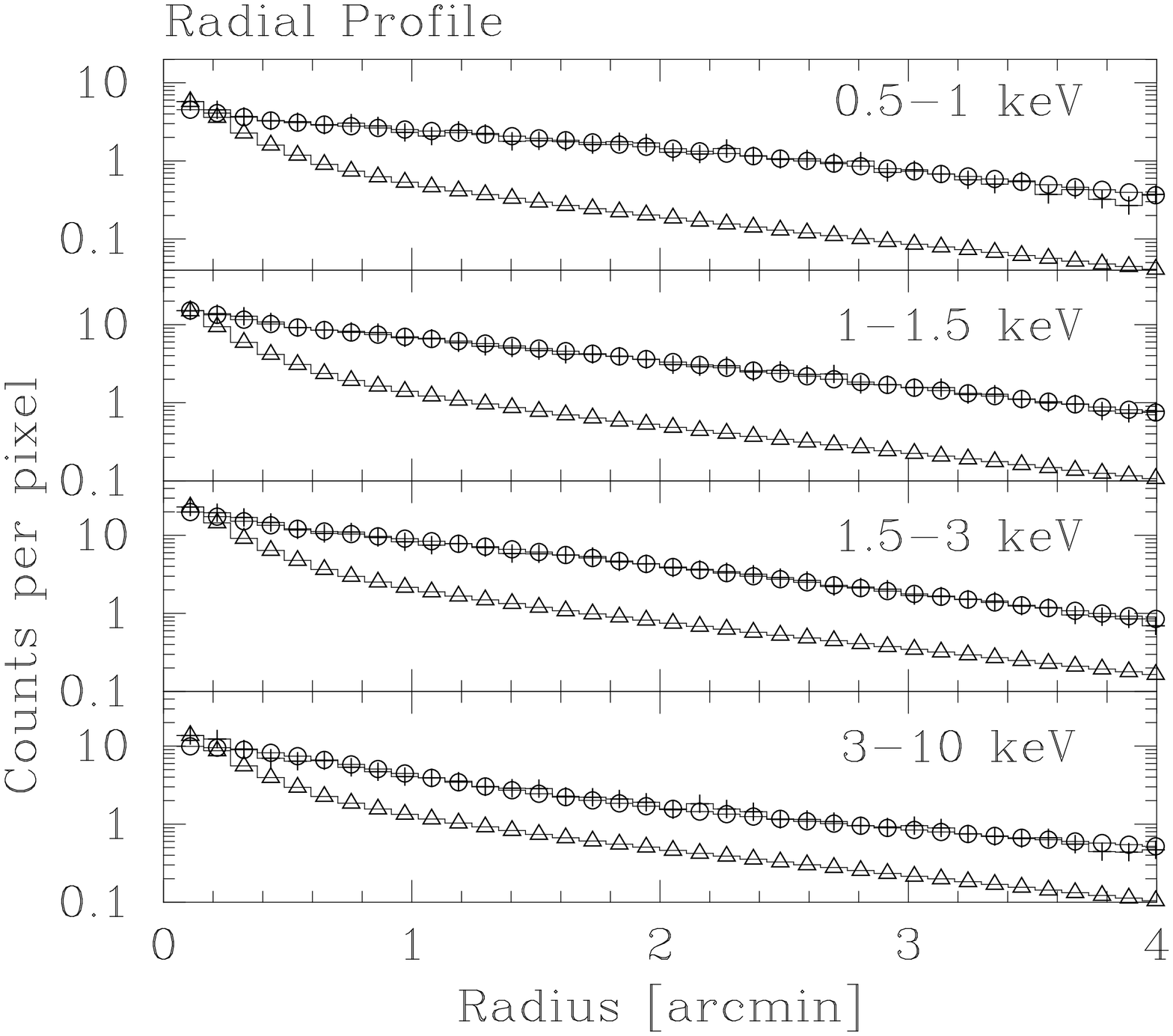,height=10.0cm,angle=270}
\caption{}
\label{radial_profile}
\end{figure}

\begin{figure}[p]
\psfig{figure=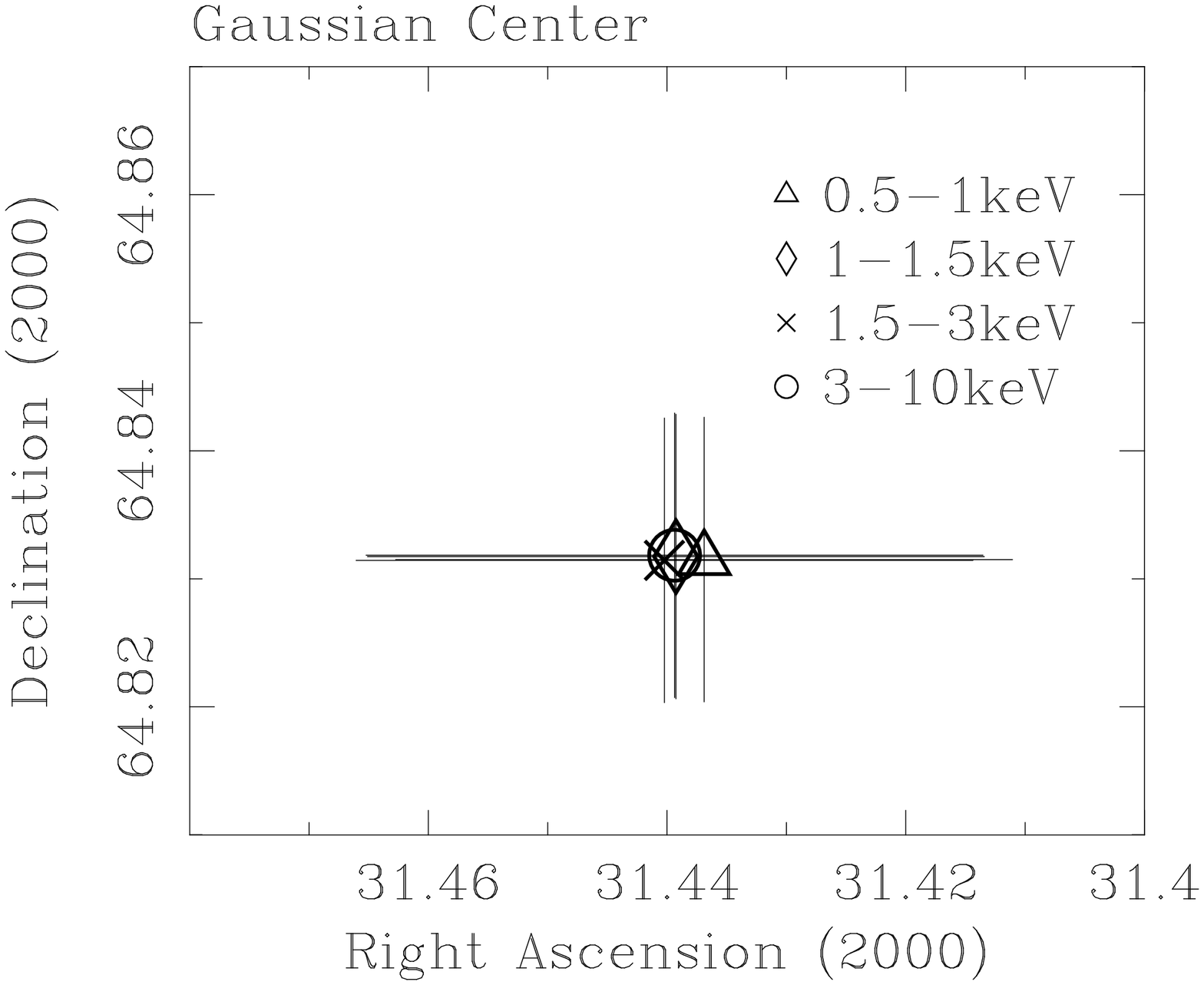,height=10.0cm,angle=270}
\caption{}
\label{xy}
\end{figure}

\begin{figure}[p]
\psfig{figure=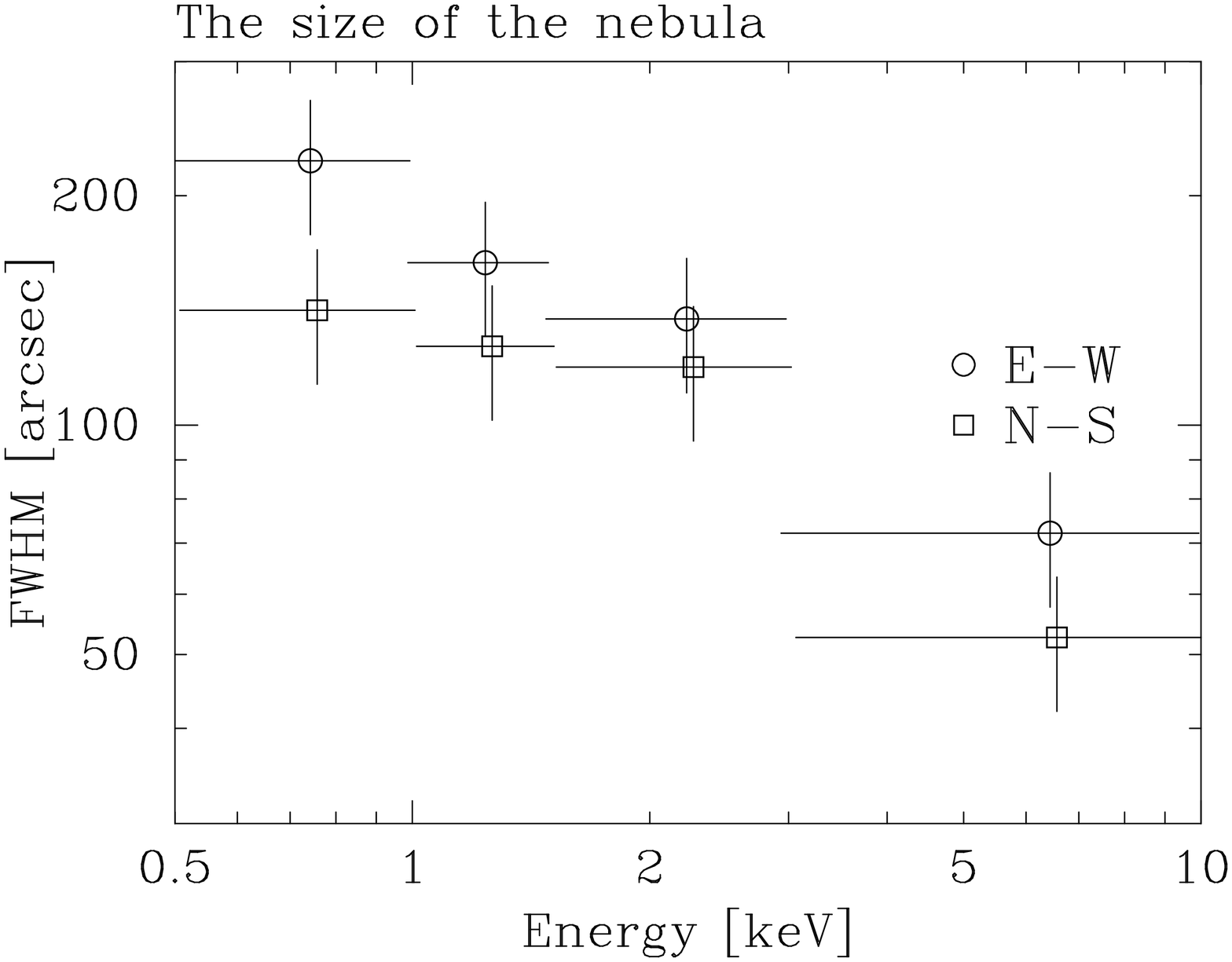,height=10.0cm,angle=270}
\caption{}
\label{sigma}
\end{figure}

	Using the multi-wavelength spectrum of 3C58 
shown in figure \ref{multi_wavelength}, we may investigate the
time dependent energy loss properties of the nebula. 
In the following discussions we
use the relation $F_{{\rm X}}{\rm [Jy]} = 10^{13.5} \cdot \nu{\rm
[Hz]}^{-1.1}$ derived from the best-fit spectral parameters of
power-law plus blackbody model. This relation is shown by a dotted
line in figure \ref{multi_wavelength}. For comparison between the Crab 
nebula and 3C58, we have summarized the basic parameters of these two
objects in table \ref{basic}.

	As noted by Green and Scheuer (1992) (hereafter GS), a sharp
break must exist around $ 5\times 10^{10}$ Hz by comparing the radio
data with the infrared data, below which the relation $F_{{\rm R}}{\rm
[Jy]} = 10^{2.3} \cdot \nu{\rm[Hz]}^{-0.09}$ is applicable. 
Extrapolation of the X-ray spectrum back toward lower energies yields a flux
density below the upper limits from IRAS data (Figure \ref{multi_wavelength}). However, 
standard synchrotron loss models for a power law distribution of electrons
predict a radiated spectrum whose index beyond the break is 0.5 larger than
the radio spectral index of $\sim 0.1.$ The fact that the X-ray energy
index is much larger than $0.6$ clearly rules out a single break under
such a scenario. 

	Based on the energy loss model of Pacini and Salvati (1973),
Bandiera et al. (1996) proposed that the low frequency spectral break 
for the plerionic
component of the SNR G11.2--0.3 be identified with $\nu_{{\rm c}}$, 
a break associated with the
radiative lifetime at $\tau$, a characteristic slowing-down time scale
of the pulsar, rather than $\nu_{{\rm b}}$, the break frequency
associated with the radiative lifetime at the current time $t$
(Pacini, Salvati 1973; Reynolds, Chanan 1984; Bandiera et al. 1996;
Woltjer et al. 1997). If the break at $\nu \simeq 50$~GHz for 3C58 is due to
the radiative lifetime at the current time, an unreasonably large
magnetic field of $\stackrel{>}{_\sim} 3\times 10^{-3}$~G has to exist in the nebula. We thus interpret
this break as due to the decreasing energy output from the pulsar. 
However, a simple model does
not reproduce the whole electromagnetic spectrum from radio through
gamma-ray, and fine tuning of pulsar's braking law and the nebular
expansion law seems to be needed (e.g., Reynolds, Chanan 1984).
Further, if a second spectral break is invoked, it would still need
to be at frequencies below $\sim 10^{12}$~Hz in order to be compatible
with the IR upper limits. Woltjer et al. (1997) have shown that the
sharpness of these breaks cannot be reproduced by any standard evolutionary
scenario with a fixed braking index. Instead, their results support the
suggestion by GS that a sudden reduction in the particle injection rate is
required at some point in time.

Difficulties in reproducing the detailed spectra of plerionic remnants 
based on a single power-law particle distribution are well-known
(Reynolds, Chevalier 1984; Reynolds, Chanan 1984; Atoyan, Aharonian
1996; Kennel, Coroniti 1984a, b).  In addition to an inherent change
in the particle injection spectrum, there are other possible mechanisms
which can lead to spectral breaks. In particular, the presence of
multiple particle populations with different spectral properties
can lead to such features. Such a scenario has some observational
support. F${\rm {\ddot u}}$rst et al. (1988)
observed the Crab-like SNR G21.5--0.9 at 22.3 GHz and compared the
result with the Einstein HRI map previously obtained by Becker and
Szymkowiak (1981); they noted that
maximum X-ray emission coincides with a minimum in the radio emission. 
Recent {\it Chandra} observations of G21.5--0.9 (Slane et al. 2000)
confirm the bright compact central X-ray emission, thus indicating
a difference in the spatial distribution of at least some of the
radio and X-ray producing particles. 
These results seem to suggest complex injection processes for
radio and X-ray producing particles. However, we can not yet say whether
this scenario holds for 3C58 with the currently available data.

\subsection{Central Object}

	Observationally, a variety of X-ray spectra and pulse profiles
have been reported for known rotation powered pulsars. Some show small
duty cycle pulses with large pulsed fractions while others show
sinusoidal pulse profiles with small pulsed fractions (e.g., ${\rm
\ddot{O}}$gelman et al. 1993; Finley et al. 1992; Halpern, Ruderman
1993). Also, pulsed fractions are known to vary with photon energy.

	If the X-ray spectrum of the pulsed component of the compact
object in 3C58 is the same as that of the blackbody component determined from
the spectral analysis, upper limits for the pulsed fraction of the
compact source are obtained as 62--82\% in the 0.7--10 keV band
depending on the frequency range as described above. On the other
hand, if the pulsed emission is hard, we may expect the number
fraction of photons from the compact source to increase with increasing
energy. From the simultaneous fit of all the SIS and GIS data using
two power-law models modified by the interstellar
absorption, of which one component has a photon index fixed at that
for the Crab pulsar ($\gamma = 1.7$; Lyne, Graham-Smith 1990), 
we obtained an upper limit of 13\% of the total counts
for the number fraction from this
component. Using this value, upper limits
for the pulsed fraction for the compact source, in the case of the
flat pulsar spectrum, are obtained as 
56\%, 74\%, 72\%, and 67\% in the 0.01--1, 1--8, 8--32, and 32--64 Hz
range, respectively, in the whole energy band (0.7--10 keV).

Helfand et al. (1995) obtained a blackbody
temperature ($3.5\times10^6 {\rm K}$) and an emitting area 
($\sim 4.5\times 10^{10} d_{{\rm3.2 kpc}}^{2} {\rm cm^2}$) for the
central compact source. They found that the temperature is higher than
that expected from any current cooling models (e.g., Nomoto, Tsuruta
1987; Page, Applegate 1992) and that the associated emitting area is
smaller than that expected from the whole neutron star
surface. Therefore, they attributed the emission to a heated polar
cap. We have confirmed the existence of a blackbody spectrum with the
temperature of $T_{{\rm blackbody}} =
(5.1\stackrel{+0.6}{_{-0.5}})\times 10^6 \;{\rm K}$ and the emitting
area of $(1.2\stackrel{+0.8}{_{-0.7}})\times 10^{10} d_{{\rm
3.2kpc}}^{2} \; {\rm cm^2}$.  

We can infer some of the properties if the neutron star by considering
the empirical relationship between the X-ray luminosity, $L_X$, of
pulsar/plerion systems and the spin-down power, $\dot E$, of the
central pulsar derived by Seward and Wang (1988) based upon
observations with the Einstein Observatory: $\log L_X = 1.39 \log \dot
E - 16.6$. Adopting this relationship for 3C58, and converting the
unabsorbed ASCA (0.5-10 keV) flux to the Einstein IPC bandpass
(0.2-2.4 keV) for which the relationship was derived, we obtain $\dot
E \sim 4 \times 10^{36} d_{\rm 3.2kpc}^2 {\rm\ ergs\ s}^{-1}$.
This value of $\dot E$ is two orders of magnitude smaller than
that for the Crab pulsar, clearly indicating that not all young
neutron stars are ``Crab-like'' in terms of their spin-down
characteristics.

Assuming that the pulsar has spun down significantly from
its initial period and has lost its rotational energy at
the rate of the pure magnetic dipole radiation, the SNR age (814 yr
assuming an association with SN 1181) and energy loss rate can be
used to estimate the initial spin period (Seward and Wang 1988):

\begin{equation}
P = \left(\frac{2 \pi^2 I}{\dot E t}\right)^{1/2} \approx
0.44 d_{3.2}^{-0.7} {\rm\ s}.
\end{equation}

The spin-down rate is then $\dot P \sim 8.5 \times 10^{-12} d_{3.2}^{-1} {\rm\
s\ s}^{-1}$ and the surface magnetic field strength is
$B \approx 6 \times 10^{13} d_{3.2}^{-0.7} {\rm\ G},$
which is at the high end of magnetic field values for the population
of young pulsars.
We note that variations as large as a factor of 10 or so
are observed in the $L_x$ vs. $\dot E$ relationship, which could result
in an overestimate of the field strength of more than a factor of 3.
More importantly, relaxing the assumption that the initial spin period
was much shorter than the current period leads to a broad range of
permissible magnetic fields. 

If the emitting area inferred from the blackbody component described
above corresponds to a standard dipole-geometry polar cap, we
may follow another path toward determining the spin properties of
the central object (Helfand et
al. 1995, equation (4); Lyne, Graham-Smith 1990, equation (14.7))

\begin{equation}
P = 58 [{\rm ms}] (\frac{r_{{\rm cap}}}{{\rm 0.6km}})^{-2}
(\frac{ R_*}{{\rm 10km}})^3
\cos^2(90^{\circ} - \theta).
\label{period}
\end{equation}
The notations are basically the same as those of Helfand et
al. (1995); $r_{{\rm cap}}$ is the polar cap radius, $R_*$ is the
neutron star radius, $\theta$ is the angle between the rotation and
magnetic axes, and the units are in cgs unless otherwise
specified. The spin down rate is thus
${\dot P} = 1.1\times 10^{-12} [{\rm s \cdot s^{-1}}] 
(\frac{r_{{\rm cap}}}{{\rm 0.6km}})^{-2} (\frac{R_*}{{\rm 10km}})^3 
\cos^2(90^{\circ} - \theta),$ and the
current spin down power of the pulsar is 
${\dot E} 
= 2.3\times 10^{38} [{\rm ergs\cdot s^{-1}}] (\frac{I_*}{10^{45}})
(\frac{r_{{\rm cap}}}{{\rm 0.6km}})^{4}(\frac{R_*}{{\rm 10km}})^{-6}
\cos^{-4}(90^{\circ} - \theta).$

The conversion rate of the spin down power to the power-law X-ray
luminosity is
\begin{equation}
\frac{L_{{\rm X}}}{{\dot E}} = 9\times 10^{-5}
(\frac{L_{{\rm X,pow}}}{2.0\times 10^{34}})(\frac{I_*}{10^{45}})^{-1}
(\frac{r_{{\rm cap}}}{{\rm 0.6km}})^{-4}(\frac{R_*}{{\rm 10km}})^{6}
\cos^{4}(90^{\circ} - \theta) 
\label{ratio1}
\end{equation}
which is anomalously low compared to other young objects (Seward, Wang 
1988; Becker, Tr${\rm \ddot u}$mper 1997; Kawai, Tamura, Shibata 1998). 
This again leads to a relatively strong magnetic field ($\sim 10^{13}$~G),
but we note that the $\dot E$ result depends very strongly on the inferred
polar cap size, which is quite uncertain. In particular,
one of the uncertainties is the
spectral fit by a simple blackbody model. More realistic models have
been constructed and applied to several neutron star candidates (e.g.,
Zavlin, et al. 1996; Zavlin, et al. 1998). These models give a factor
of $\simeq 2$ lower temperature and a factor of $\stackrel{>}{_\sim} 4$ larger
emitting radius, values which are still too hot and too small to
reconcile with emission from the entire surface of a neutron star. 
In the discussion below, we adopt the value of $\dot E$ derived from
the Seward \& Wang (1988) relationship under the assumption that
the spin-down power of the source in 3C58 is similar to that of other
objects powering synchrotron nebulae.

\subsection{Energy-Size Relation and the Particle Flow in the Nebula}

The measured size of 3C58, and its dependence on energy, can be used
to derive constraints on the particle flow in the nebula. The overall
picture is as follows (see, e.g., KCa). A central pulsar injects
energy into the nebula at a rate $\dot E$ in the form of a
relativistic particle wind. The wind zone is bounded by an MHD shock
at a radius $r_s$ beyond which the bulk particle flow is decelerated and
the pressure is increased. A non-relativistic flow transports plasma
and magnetic flux from the shock region to the edge of the nebula.
Synchrotron emission from decelerated particles in a primarily
toroidal magnetic field forms the observed nebula. Equipartition
pressure of the magnetic field and the particle energy density drives
the nebular expansion and determines the nebular structure. In this
picture, an important parameter which characterizes the system is the
magnetization parameter, $\sigma$, the ratio of the Poynting to
particle energy fluxes in the upstream wind. The parameter can be
constrained by applying pressure and flow boundary conditions at the
edge of the nebula.

From radio observations (GS), the equipartition pressure in the
nebula is estimated to be $P_{\rm neb} \sim 10^{-10} {\rm\ dyn\
cm}^{-2}$, which is also two orders of magnitude smaller than in the
Crab nebula. The corresponding magnetic field of the nebula is $\sim 5
\times 10^{-5}$~G.  Balancing the ram pressure of the wind, $\dot E/(4
\pi c R_s^2)$, with the internal pressure of the nebula, we estimate
that the distance from the pulsar to the confinement shock region is
$R_s \sim 0.1$~pc. This corresponds to an angular size of $\sim
6.5$~arcsec. Using deep VLA imaging, Frail and Moffett (1993)
discovered an elongated radio wisp just west of the compact X-ray
source, which they argue is the termination shock from the pulsar
wind. Using the ROSAT HRI position determination for the X-ray source
(Helfand et al. 1995), the wisp lies at a distance of $\sim
6.7$~arcsec (0.1 pc) from the compact source, in excellent agreement
with the above estimates.  We note that Frail and Moffett (1993) quote
an angular separation of 2.6 arcsec based upon the position
determination of the compact source using Einstein HRI data; this
difference is consistent with the ROSAT result considering the aspect
determination errors of the two satellites. The radio wisp thus lies
at a distance consistent with the estimated shock position and appears
to represent a distinct signature of the termination shock.

From equation 5.15 of KCa, the magnetization parameter of the wind is
related to the velocity profile of the nebula by $v(z) \approx 3
\sigma [1 + (3 \sigma z^2)^{-1/3}]$ where $z \equiv r/r_s$. The
apparent size of the radio nebula ($10^\prime.3 \times 6^\prime.3$)
yields a nebular size $r_N = (4.4 - 2.9) d_{3.2}$~pc. It is
interesting to note that this yields $r_N/r_s \approx 15 - 100$ which
is similar to that for the Crab ($\sim 20$). Using the age of 814 yr
thus gives a mean expansion velocity of $(5.4 - 3.5) \times 10^8
d_{3.2} {\rm\ cm\ s}^{-1}$. Assuming homologous expansion, the current
velocity is 2/5 of the averaged value. Thus, using $z_N = r_N/r_s$ we
find $\sigma \sim (2 - 6) \times 10^{-3}$, comparable to that for the
Crab nebula.  Alternatively, constant velocity of expansion leads to
$\sigma \sim (6 - 15) \times 10^{-3}$. Thus, the prime assumption of
the KC model, namely a small magnetization parameter corresponding to
a particle-dominated wind, is valid here.

Well beyond the wind termination shock, the magnetic field in the nebula is
primarily the wound-up field from the pulsar. The radial profile of this
toroidal field is $B(r) \propto [r v(r)]^{-1}$ (Aschenbach, Brinkmann
1975; Ku et al. 1976).
Under the assumption of a power law velocity profile $v(r) \propto
r^{-m}$ (Aschenbach, Brinkmann 1975), the magnetic field strength
then varies as $B(r) \propto r^{m-1}$. Due to the finite synchrotron
lifetime of the emitting electrons, the measured nebular radius is a
function of the observed photon energy: $r \propto E^{-1/(6m-2)}$
(Aschenbach and Brinkmann 1975).  Using the measured energy dependence
of the nebular size summarized in Section 3.3, we find $m =
0.7^{+0.2}_{-0.1}$ for both EW and NS directions. The magnetic field
strength thus decreases slowly with radius. For the Crab nebula, Ku et
al. (1976) find an energy-size relationship which leads to $m =
1.46^{+0.10}_{-0.08}$. The velocity law is thus flatter for 3C58,
indicating a smaller deceleration of the particle flow in the nebula
which corresponds with the smaller inferred confinement pressure for
3C58 relative to that for the Crab.

From the above discussion, we propose a semi-quantitative picture of
3C58 as follows. The current spin-down power of the central pulsar in
3C58 is $\dot E \sim 4 \times 10^{36} {\rm\ erg\ s}^{-1}$ which is two
orders of magnitude smaller than that of the Crab pulsar. From radio
observations, the equipartition pressure in the nebula is $\sim
10^{-10}{\rm\ dyn\ cm}^{-2}$ which is also about two orders of
magnitude smaller than for the Crab nebula.  Consequently, the
position of the standing shock where the ram pressure of the pulsar
wind balances the nebular pressure is at about the same distance from
the pulsar, $r_s \sim 0.1$~pc, for both 3C58 and the Crab. The
magnetization parameter $\sigma \sim (2 - 15) \times 10^{-3}$ is of
roughly the same order as that for the Crab, indicating a
particle-dominated flow at the point of injection. Using these
parameters, we obtain the upstream magnetic field (KCb, equation 2.2)
as $B_1 = 2.6 \times 10^{-6} (\dot E/(4 \times 10^{36} {\rm\ erg\
s}^{-1}))^{1/2} (\sigma/(5 \times 10^{-3}))^{1/2} (r_s/0.1{\rm\
pc})^{-1}$~G. The downstream magnetic field is amplified by a factor
of 3 for the small $\sigma$ limit (KCa, equation 4.15d) and $B_2 \sim
7.8 \times 10^{-6}$~G. The magnetic field in the inner regions
increases with increasing radius, and reaches the equipartition value
of $\sim 5 \times 10^{-5}$~G at a radius $\bar r = r_s \bar z = r_s (3
\sigma)^{-1/2} \sim 0.8$~pc. If we approximate the bulk motion
velocity in the nebula by a power law, we find a flatter profile than
for the Crab, indicating a slowly decreasing magnetic field strength
in the outer nebula and a smaller deceleration with lower confinement
pressure in 3C58 than for the Crab.

\section{Conclusion}

With the broad band spectroscopic capability of ASCA, we have obtained
accurate X-ray spectra of the SNR 3C58. Using both the SIS and GIS
spectra, we find the integrated emission to be best described by a
power law with $\gamma = 2.1 \pm 0.1$ accompanied by a blackbody with
$T_{\rm blackbody} = (5.1^{+0.6}_{-0.5}) \times 10^6$~K and an
emitting area of $(1.2^{+0.8}_{-0.7}) \times 10^{10} d_{3.2}^2 {\rm\
cm}^2$. The column density is $N_H = (3.3 \pm 0.4) \times 10^{21}{\rm\
cm}^{-2}$. We associate the blackbody component with the central
compact X-ray source in the remnant. No pulsations are detected from
the central emission in 3C58. We set upper limits of $\sim 55-80\%$
for the pulsed fraction of the compact source, depending upon the
spectral characteristics of any pulsed emission.

Using images of the nebula in different energy bands, we have
established the spectral softening with radius expected for a
synchrotron spectrum injected from a central source. We have used
these results to obtain, for the first time, the energy-size relation
for the 3C58 nebula. By comparing this result to model of Aschenbach
and Brinkmann (1975) we find the bulk motion velocity and the magnetic
field strength to be flatter functions of radius than for the Crab
nebula.  In addition, we have used the results to derive constraints
on the interior pressure and magnetic field, as well on the shock
magnetization parameter, using the models of KC. Small confinement
pressure and the small deceleration of expansion of the 3C58 nebula
has been suggested for the 3C58 nebula compared to the Crab.

\vspace{2cm}

	The authors are grateful to the referee, Dr. Robert Petre, for invaluable
comments and suggestions, which largely improved the draft. The
authors are grateful to all the members of the ASCA team. KT was
supported in part by Research Fellowships of the Japan Society for the
Promotion of Science for Young Scientists. POS is supported in part by
NASA Contract NAS8-39073 and Grants NAG5-2638 and NAG5-3486.

\def\baselinestretch{1.0}

\setlength{\parindent}{0.0cm}

\newpage
\section*{References}
\small

Anders E., Grevesse N. 1989, Geochimica et Cosmochimica Acta 53, 197 

Asaoka I., Koyama K. 1990, PASJ 42, 625

Aschenbach B., Brinkmann W. 1975, A\&A 41, 147

Atoyan A.M., Aharonian F.A. 1996, MNRAS 278, 525

Bandiera R., Pacini F., Salvati M. 1996, ApJL 465, L39

Becker R.H., Szymkowiak A.E. 1981, ApJL 248, L23

Becker R.H., Helfand D.J., Szymkowiak A.E. 1982, ApJ 255, 557

Becker W.,  Tr${\rm ddot u}$mper J. 1997, A\&A 326, 682

Bietenholz M.F., Frail D.A., Hankins T.H. 1991, ApJL 376, L41

Burke B.E., Mountain R.W., Daniels P.J., Cooper M.J., Dolat V.S. 1994, IEEE Trans. Nuclear Science 41, 375

Chiueh T., Li Z-Y., Begelman M.C. 1998, ApJ 505, 835

Davelaar J., Smith A., Becker R.H. 1986, ApJL 300, L59

Fichtel C.E., Bertsch D.L., Chiang J., Dingus B.L., Esposito J.A., Fierro J.M., Hartman R.C., Hunter S.D. et al. 1994, ApJS 94, 551

Finley J.P., ${\rm \ddot{O}}$gelman H., Kiziloglu U. 1992, ApJL 394, L21 

Frail D.A., Moffett D.A. 1993, ApJ 408, 637

F${\rm {\ddot u}}$rst E., Handa T. Morita K., Reich P., Reich W.,
Sofue Y. 1988, PASJ 40, 347

Goss W.M., Schwartz U.J., Wesselius P.R. 1973, A\&A 28, 305

Gotthelf E. 1996, ASCA News 4, p31

Gotthelf E., Ishibashi K. 1997, X-ray Imaging and Spectroscopy of
Cosmic Hot Plasmas, ed Makino F., Mitsuda K. (Universal Academy Press,
Tokyo) p631

Green D.A., Gull S.F. 1982, Nature 299, 606

Green D.A. 1986, MNRAS 218, 533

Green D.A., Scheuer P.A.G. 1992, MNRAS 258, 833 (GS)

Green D.A. 1994, ApJS 90, 817

Halpern J.P., Ruderman M. 1993, ApJ 415, 286

Helfand D.J., Chanan G.A., Novick R. 1980, Nature 283, 24

Helfand D.J., Becker R.H., White R.L. 1995, ApJ 453, 741

Hester J.J., Scowen P.A., Sankrit R., Burrows C.J., Gallagher III J.S., Holtzman J.A., Watson A., Trauger J.T. et al. 1995, ApJ 448, 240

Kawai, N., Tamura, K., Shibata, S. 1998, Neutron Stars and
Pulsars Thirty Years after the Discovery, ed. N. Shibazaki, N. Kawai,
S. Shibata, \& T. Kifune (Tokyo: Universal Academy Press), 449

Kennel C.F., Coroniti F.V. 1984, ApJ 283, 694 (KCa)

Kennel C.F., Coroniti F.V. 1984, ApJ 283, 710 (KCb)

Ku W., Kestenbaum H.L., Novick R., Wolf R.S. 1976, ApJL 204, L77

Liedahl D.A., Osterheld A.L., Goldstein W.H. 1995, ApJL 438, L115

Lyne A.G., Graham-Smith F. 1990, Pulsar Astronomy (Cambridge
University Press, Cambridge)

Makishima K., Tashiro M., Ebisawa K., Ezawa H., Fukazawa Y., Gunji S.,
Hirayama M., Idesawa E. et al. 1996, PASJ 48, 171

Mewe R., Gronenschild E.H.B.M., van den Oord G.H.J. 1985, A\&AS 62, 197

Morrison R., McCammon D. 1983, ApJ 270, 119

Nomoto K., Tsuruta S. 1987, ApJ 312, 711

Ohashi T., Ebisawa K., Fukazawa Y., Hiyoshi K., Horii M., Ikebe Y., Ikeda H., Inoue H. et al. 1996, PASJ 48, 157

${\rm \ddot{O}}$gelman H., Finley J.P., Zimmerman H.U. 1993, Nature 361, 136

Pacini F., Salvati M. 1973, ApJ 186, 249

Page D., Applegate J.H. 1992, ApJL 394, L17

Panagia N., Weiler K.W. 1980, A\&A 82, 389

Rees M.J., Gunn J.E. 1974, MNRAS 167, 1

Reynolds S.P., Chevalier R.A. 1984, ApJ 278, 630

Reynolds S.P., Chanan G.A. 1984, ApJ 281, 673

Reynolds S.P., Aller H.D. 1985, AJ 90, 2312

Reynolds S.P., Aller H.D. 1988, ApJ 327, 845

Roberts D.A., Goss W.M., Kalberla P.M.W., Herbstmeir U., Schwarz U.J. 
1993, A\&A 274, 427

Saken J.M., Fesen R.A., Shull J.M. 1992, ApJS 81, 715

Salter C.J., Reynolds S.P., Hogg D.E., Payne J.M., Rhodes P.J. 1989, ApJ 
338, 171

Sedov L.I. 1993, Similarity and dimensional methods in mechanics 10th
edition (CRC Press, Boca Raton)

Serlemitsos P.J., Jalota L., Soong Y., Kunieda H., Tawara Y., Tsusaka
Y., Suzuki H., Sakima Y. et al. 1995, PASJ 47, 105

Seward F.D., Harnden F.R., Helfand D.J. 1984, ApJL 287, L19

Seward F.D., Wang Z.R. 1988, ApJ 332, 199

Slane P. 1994, ApJ 437, 458

Slane P., Lloyd N. 1995, ApJL 452, L115

Slane P., Chen Y., Schulz N.S., Seward F.D., Hughes J.P.,
Gaensler B.M. 2000, ApJL in press

Tanaka Y., Inoue H., Holt S.S. 1994, PASJ 46, L37

Toor A., Seward F.D. 1974, AJ 79, 995

Torii K., Tsunemi H., Dotani T., Mitsuda K. 1997, ApJL 489, L145

Vasisht G., Aoki T., Dotani T., Kulkarni S.R., Nagase F. 1996, ApJL 456, L59

Vaughan B.A., van der Klis M., Wood K.S., Norris J.P., Hertz P.,
Michelson P.F., van Paradijs J., Lewin W.H.G. et al. 1994, ApJ 435, 362

Weiler K.W., Panagia N. 1978, A\&A 70, 419

Williams D.R.W. 1973, A\&A 28, 309

Woltjer L., Salvati M., Pacini F., Bandiera R. 1997, A\&A 325, 295

Wu C-C. 1981, ApJ 245, 581

Yancopoulos S., Hamilton T.T., Helfand D.J. 1994, ApJ 429, 832

Zavlin V.E., Pavlov G.G., Shibanov Yu.A. 1996, A\&A 315, 141

Zavlin V.E., Pavlov G.G., Tr${\rm {\ddot u}}$mper J. 1998, A\&A 331, 821

\onecolumn

\begin{table}[p]
\begin{center}
\caption{Observation date and mode.}
\label{observation_date_mode}
\begin{tabular}{c c c c c c c} \hline\hline
Number & Start Time (UT) & End Time (UT) & SIS0 & SIS1 & GIS2 & GIS3 \\ \hline 
1 & 14:13 Sep. 12, 1995 & 07:50 Sep. 13, 1995 & C1 & C3 & PH nominal & PH nominal \\ 
2 & 07:05 Sep. 13, 1995 & 17:20 Sep. 13, 1995 & C0 & C2 & PH timing & PH timing \\ 
3 & 17:21 Sep. 13, 1995 & 04:00 Sep. 14, 1995 & C2 & C0 & PH timing & PH timing \\
4 & 04:01 Sep. 14, 1995 & 14:00 Sep. 14, 1995 & C1 & C3 & PH timing & PH timing \\
\hline
\end{tabular}
\end{center}
\end{table}

\begin{table}[p]
\begin{center}
\caption{Effective exposure time [ks].}
\label{exposure}
\begin{tabular}{c c c c c} \hline\hline
Number & SIS0 & SIS1 & GIS2 & GIS3 \\ \hline
1 & 22.05 & 22.19 & 27.14 & 27.05 \\
2 & 12.31 & 12.31 & 15.28 & 15.37 \\
3 & 15.24 & 15.34 & 18.08 & 18.08 \\
4 & 13.91 & 13.93 & 16.98 & 16.96 \\ \hline
\end{tabular}
\end{center}
\end{table}

\begin{table}[p]
\begin{center}
\caption{Spectral parameters of the whole nebula with the 90\%
confidence errors. Photon index ($\gamma$), normalization (${\rm [photons\cdot
s^{-1}\cdot keV^{-1}\cdot cm^{-2}]}$ at 1 keV), and the absorbing
column density ($n_{{\rm H}} {\rm [10^{21}cm^{-2}]}$) are summarized.}
\label{all_spectra_parameters}
\begin{tabular}{ c c c c c } \hline \hline
 & S0C1 + S1C3 & S0C0 + S1C2 & S0C2 + S1C0 & G2 + G3\\ \hline
$\gamma$ & $2.24 \pm 0.05$ & $2.21 \pm 0.08 $ & $2.26\stackrel{+0.07}{_{-0.06}}$ & $2.32\pm 0.07 $ \\
Normalization & $(4.9\stackrel{+0.3}{_{-0.2}})\times10^{-3}$ & $(4.6\pm 0.4)\times10^{-3}$ & $(4.7\pm 0.3)\times10^{-3}$ & $(4.7\pm 0.4)\times 10^{-3}$ \\
$n_{\rm H}$ & $3.7\pm 0.2$ & $4.0\pm 0.4$ & $4.1\pm 0.3 $ & $3.5 \pm 0.5 $ \\ \hline
$ \chi ^2 /{\rm d.o.f.}$ & 302.1/337 & 289.2/280 & 354.1/315 & 436.3/396 \\ \hline
\end{tabular}
\end{center}
\end{table}

\begin{table}[p]
\begin{center}
\caption{
99\% confidence upper limits of pulsed fraction in \%.}
\label{upper_limit_new}
\begin{tabular}{ c c c c } \hline \hline
                     & \multicolumn{3}{c}{Energy Range [keV]} \\ 
Frequency Range [Hz] & 0.7--10 	& 0.7--2 & 2--10  \\ \hline
0.01--1		     & 7    	& 12    & 11  	\\
1--8                  & 10   	& 14    & 9   	\\ 
8--32                 & 9    	& 19    & 15    \\ 
32--64                & 9    	& 14    & 12    \\ \hline
\end{tabular}
\end{center}
\end{table}

\begin{table}[p]
\begin{center}
\caption{Basic parameters of the Crab nebula and 3C58.}
\label{basic}
\begin{tabular}{l c c } \hline \hline
  & Crab & 3C58 \\ \hline
Distance [kpc] & 2 & 3.2  \\
Age [yr] in 1995 & 941 & 814 \\
Size [pc] $^a$ & 1.2 & $2.9\times 2.2$\\
$L_X$ [${\rm ergs\,s^{-1}}$] $^b$& $2.1\times 10^{37}$& $2.4\times 10^{34}$ \\
$\nu_B$ [Hz] $^c$& $1\times 10^{13}$ & $5\times 10^{10}$\\ \hline
\end{tabular}
\end{center}
Notes. --- $^a$ The size of the nebula at 1 keV defined by the
Gaussian FWHM. The value for the Crab nebula is based on Ku et al. (1976).\\
$^b$ Total (nebula plus compact source) X-ray luminosity in the energy range 0.5--10 keV. The value for
the Crab nebula is based on Toor and Seward (1974).\\
$^c$ Frequency of the spectral break. The values are adopted from
Green (1994) and the references therein.\\
\end{table}

\section*{Figure captions}

Figure 1: Representative energy spectra of the whole SNR obtained
by S0C1 and GIS2. The solid lines in the spectra show the best fit
power-law model modified by the interstellar absorption. The lower
panels show the residuals.\\

Figure 2: Upper panel shows the upper limits of the emission measure
of the thermal component of MEKAL model. Lower panel shows the upper
limits of the density for the filling factor of unity (dashed line),
and the filling factor of 1/4 (solid line) which corresponds to the
Sedov type self similar solution.\\

Figure 3: Multi-wavelength spectra of the Crab nebula (dashed line) and
3C58. The dotted line shows the extrapolation of the X-ray spectrum.\\

Figure 4: Photon index of the power-law component as a function of
radius with 90\% confidence errors. Squares show the case without the
blackbody component and the circles show the case with the blackbody
component.\\

Figure 5: Radial profile of the image, the best-fit model function and
the PSF as the results of the fitting. Crosses show the data points,
circles show the best-fit model function and the triangles show the
PSF.\\

Figure 6: Position of the Gaussian center, $x_0$ and $y_0$ in
equatorial coordinate (Epoch 2000).\\

Figure 7: The size of the nebula as a function of the observing
energy.\\

\clearpage

\clearpage

\clearpage

\clearpage

\clearpage

\clearpage

\clearpage

\end{document}